\documentclass[prl, etal, twocolumn,  longbibliography, amsmath,superscriptaddress,amssymb]{revtex4-2}

\usepackage{graphicx}
\usepackage{subfigure}
\usepackage{emf}
\usepackage{dcolumn}
\usepackage{bm}
\usepackage{verbatim}
\usepackage{xcolor}
\usepackage{braket}
\usepackage{tikz}
\usetikzlibrary{shapes.geometric, arrows}
\usepackage{pgfplots}
\usepackage{mathtools}

\usepackage[left,pagewise]{lineno}

\begin{document}

\title{Laser plasma accelerated ultra-intense electron beam for efficiently exciting nuclear isomers}

\def\fdu {Key Laboratory of Nuclear Physics and Ion-beam Application (MoE), Institute of Modern Physics, Fudan University, Shanghai 200433,  China}
\def\sjtu {Key Laboratory of Laser Plasma (MoE), School of Physics and Astronomy, Shanghai Jiao Tong University, Shanghai 200240, China}
\def\ifsa {IFSA Collaborative Innovation Center, Shanghai Jiao Tong University, Shanghai 200240,  China}
\def\SARI{Shanghai  Advanced Research Institute, Chinese Academy of Sciences, Shanghai 201210,  China}
\def\IOP{Laboratory of Optical Physics, Institute of Physics, Chinese Academy of Sciences, Beijing 100190, China}
\def\UCA{School of Physical Sciences, University of Chinese Academy of Sciences, Beijing 100049,  China}
\def\IHEP{Institute of High Energy Physics, Chinese Academy of Sciences, Beijing 100049, China}

\author{Jie Feng}      \affiliation{\sjtu} \affiliation{\ifsa}
\author{YaoJun Li} 	\affiliation{\sjtu}\affiliation{\ifsa}
\author{JunHao Tan} 	\affiliation{\sjtu}\affiliation{\ifsa}
\author{WenZhao Wang} \affiliation{\sjtu}\affiliation{\ifsa}
\author{YiFei Li} 	\affiliation{\IOP}
\author{XiaoPeng Zhang} \affiliation{\IHEP}
\author{Yue Meng}   \affiliation{\sjtu}
\author{XuLei Ge}   \affiliation{\sjtu}\affiliation{\ifsa}
\author{Feng Liu}   \affiliation{\sjtu}\affiliation{\ifsa}
\author{WenChao Yan}\affiliation{\sjtu}\affiliation{\ifsa}
\author{ChangBo Fu} \affiliation{\fdu}
\author{LiMing Chen} \email[Corresponding author:] {lmchen@sjtu.edu.cn}\affiliation{\sjtu}\affiliation{\ifsa}
\author{Jie Zhang} \email[Corresponding author:] {jzhang1@sjtu.edu.cn} \affiliation{\sjtu}\affiliation{\ifsa}\affiliation{\IOP}


\date{\today}

\begin{abstract}

Utilizing laser plasma wakefield to accelerate ultra-high charge electron beam is critical for many pioneering applications, for example to efficiently produce nuclear isomers with short lifetimes which may be widely used. However, because of the beam loading effect, electron charge in a single plasma bubble is limited in level of hundreds picocoulomb. Here, we experimentally present that a hundred kilo-ampere, twenty nanocoulomb, tens of MeV collimated electron beam is produced from a chain of wakefield acceleration, via a tightly focused intense laser pulse transversely matched in dense plasma. This ultra-intense electron beam ascribes to a novel efficient injection that the nitrogen atom inner shell electrons are ionized and continuously injected into multiple plasma bubbles. This intense electron beam has been utilized to exciting nuclear isomers with an ultra-high peak efficiency of $1.76\times10^{15}$ particles/s via photonuclear reactions. This efficient production method of isomers can be widely used for pumping isotopes with excited state lifetimes down to picosecond, which is benefit for deep understanding nuclear transition mechanisms and stimulating gamma-ray lasers.
\end{abstract}
\maketitle

\maketitle


\section{\uppercase\expandafter{\romannumeral1}. INTRODUCTION}
The laser plasma wakefield accelerators (LWFAs) have attracted significant interests in recent years\cite{ref1,ref2,ref3,ref4} due to high acceleration gradients (hundreds of GV/m) and beam current (tens of kilo-ampere), thus not only enable GeV electron accelerators reducing to a length of centimeters\cite{ref5}, but drive secondary radiation/particle sources with ultra-high brightness/flux\cite{ref6,ref7,ref8}. In bubble\cite{ref9} or blowout\cite{ref10} regime LWFAs, an ultra-short intense laser pulse excites wakefield in underdense plasma, and the laser ponderomotive force expels electrons forming ion cavities (or called bubbles) in which electrons can be accelerated. In the last decade, some breakthroughs of LWFAs have been achieved such as ultra-high stability\cite{ref11}, multi-GeV energy\cite{ref12}, femtosecond (fs) beam duration\cite{ref13}, ultra-low energy spread\cite{ref7,ref14} etc., but beam charge is limited below hundred picocoulomb (pC). 

Recently, several hundred pC electron beams have been observed in the process of double self-injection\cite{ref15,ref16} and self-truncated ionization injection\cite{ref17a,ref17,ref18,ref19} respectively, but it is still difficult to reach nanocoulomb (nC) due to the beam loading effect of bubble regime\cite{ref10,ref20,ref21}. There are also some efforts in improving beam charge by increasing laser energy and using high-Z gas\cite{ref21a,ref21b,ref21c,ref21d}, but they still do not break through the limitation of beam loading effect. While self-modulated laser wakefield acceleration (SM-LWFA)\cite{ref22, ref23, ref24, ref25} that long laser pulse overlaps with several tens of plasma waves, large number of electrons can be trapped and accelerated in every wakefield and the beam charge can be increased tremendously to tens of nC\cite{ref26, ref27}, but it requires a hundred joule class picosecond (ps) laser facility. Moreover, directional electron beams with tens of nC charge have also been produced via vacuum laser acceleration with a plasma mirror injector\cite{ref28, ref29}. Unfortunately, the beam collimation suffers from the ponderomotive force of the laser pulse in vacuum during acceleration, which results in a large divergence angle ($\sim 20^\circ$) and a hole in the beam profile\cite{ref30}. By way of contrast, a hundred nC collimated ($\sim 3^\circ$) electron beam has been acquired from laser solid interactions with deliberately induced pre-plasma\cite{ref31}. Although laser solid interaction is propitious to realize ultra-high charge, an important limited factor is the acceleration distance resulting in electron energy usually less than 10 MeV. Moreover, tens of nC electron beam with higher energy gain can also be realized in LWFA of near-critical-density plasma\cite{ref32, ref33,ref33a}, but the beam divergence angle is usually large ($\sim 15^\circ$) due to electrons interaction with laser fields, and they would also scatter in dense plasma after the laser pulse rapid depletion. In general, for laser plasma electron acceleration, it is a great challenge to realize electron beam with large charge, tight collimation and high energy at the same time.

In this work, the electron beams with charge of $\sim$ 20 nC and small divergence angle $\sim 6^\circ$ have been generated experimentally from a tens of TW tightly focused laser pulse interacting with high density gas. Particle in cell simulation shows that a novel and efficient electron injection scenario of electrons successively ionization injecting into multiple bubbles has been realized, resulting in super-high charge surpassing the scaling law\cite{ref34} of LWFA in bubble regime. The electron beams generated have appropriate energies (10s MeV) for driving photonuclear reactions and nuclear isomers production. So, these dense and energetic electron beams hit on Indium target, and a significant amount of multiple isomers have been produced with an ultra-short time duration. This opens a new path to produce nuclear isomers with an extremely high peak efficiency.

\begin{figure}[!b]
 \centering
	\includegraphics[width=0.5\textwidth]{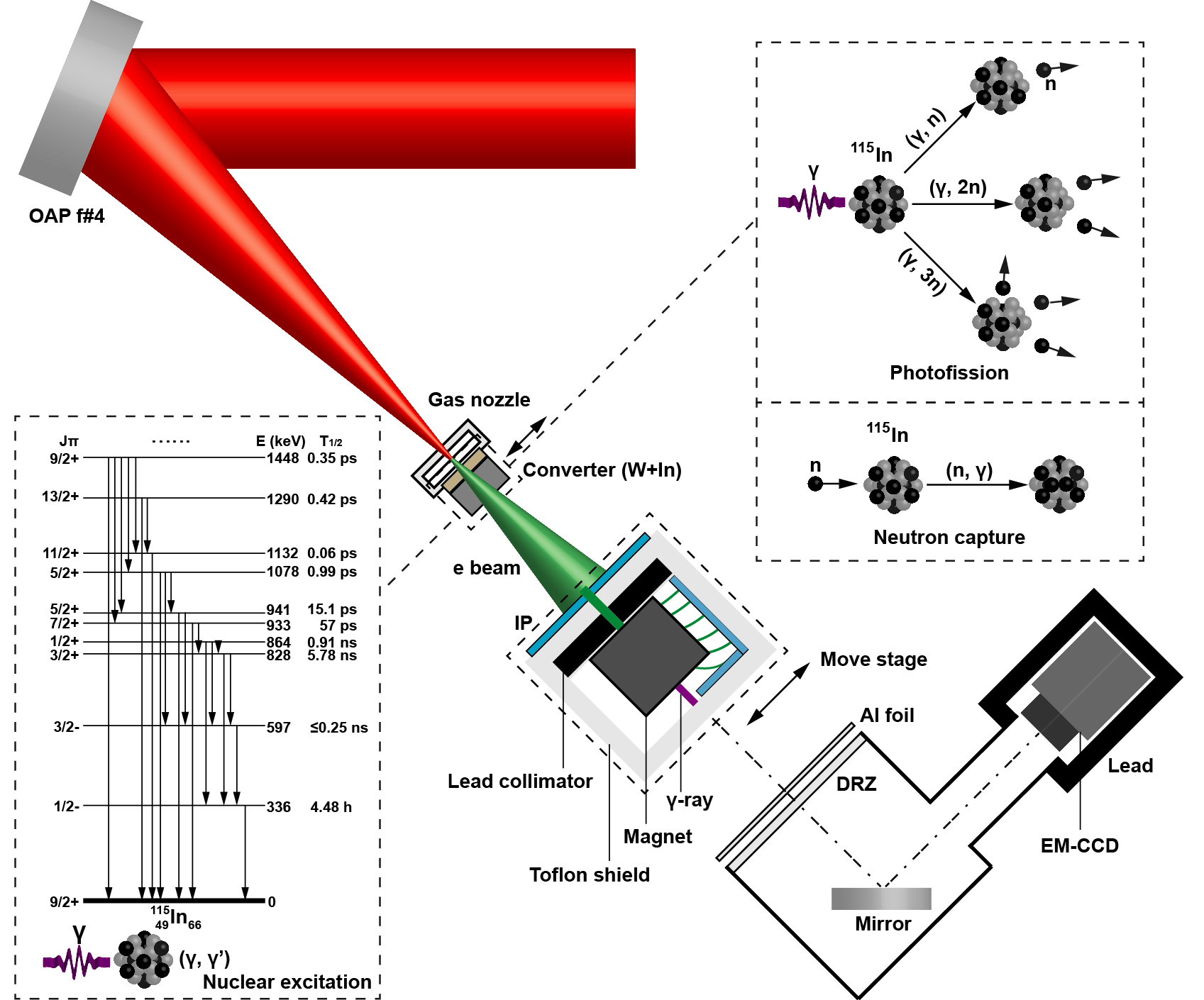}
 \caption{Experimental setup. The femtosecond laser pulse is focused onto the front edge of supersonic gas nozzle which has 10 mm length, 1.2 mm opening width. Electron beam bombards DRZ fluorescent screen (covered with 14 $\mu m$ Al film to block stray light) emitting fluorescence, which is detected by an EM-CCD to get electron beam spot image. When move in magnet, the electron beam energy spectrum is recorded by SR-type image plates (IPs). When move in converter (W+In), electron beam drives photofission reactions ($\gamma$, xn) and photonuclear excitation ($\gamma$, $\gamma'$). The generated neutrons can also induce neutron capture reaction (n, $\gamma$) in converters. Due to the converter (In) has two natural isotopes $^{115}$In (95.71\%) and $^{113}$In (4.29\%), here only shows the schematics of photofission, neutron capture and the excited states of $^{115}$In nucleus.}
\end{figure}

\section{\uppercase\expandafter{\romannumeral2}. EXPERIMENTAL SETUP}
The experiment is carried out using a Ti: Sapphire laser system at the Laboratory of Laser Plasmas of Shanghai Jiao Tong University. The experimental setup is shown in Fig. 1. A linearly polarized laser pulse with duration of 45 fs (FWHM), energy of 3.2 J is focused by an f-number 4 off-axis parabolic (OAP) mirror onto a supersonic gas jet\cite{ref35}, with well-defined uniform density distribution. The laser pulse can be focused to a main spot with radius $w_0 = 3.8 \mu m$, containing about 38\% of the laser power, i.e., $\sim$27 TW, resulting in an intensity of $\sim 5.8\times10^{19} W/cm^{2}$, corresponding normalized vector potential $a_0\approx4.9$. By regulating the gas back pressure, the outflow nitrogen gas density can range from $3\times10^{17} cm^{-3}$ to $2\times10^{19} cm^{-3}$. After the laser interaction with pure nitrogen gas, the accelerated electron beam bombards a 1 mm-thick tungsten (W) converter which is located at the downstream of nozzle to generate bremsstrahlung radiation. Then the collimated $\gamma$-ray induces the ($\gamma$,$\gamma'$) reaction, photofission ($\gamma$, n), ($\gamma$, 2n), ($\gamma$, 3n) reactions, and the followed neutron capture (n, $\gamma$) reaction in a 3 mm-thick indium (In) converter. These products of above reactions would be in nuclear isomeric states or become radionuclides, which can be identified by measuring their characteristic decay radiations with a high pure germanium detector.

\section{\uppercase\expandafter{\romannumeral3}. EXPERIMENTAL RESULTS}
\paragraph{Efficient electron acceleration.}
Laser pulse focused by a small $f$-4 OAP has intense power density but shorter Rayleigh length $l_R\approx 60 \mu m$ for $w_0 = 3.8 \mu m$. Due to the plasma bubble radius   and the laser self-focusing power $P_c\approx 17n_c/n_e [GW]$, where $\omega_{pe}=\sqrt{4\pi n_ee^2/m_e}$ is plasma frequency and $n_c$ is critical density, a higher density plasma is usually required to match the small laser focal spot ($w_0 \approx R$) for maintaining laser intensity and overcoming quick defocus\cite{ref34}. Because 27 TW focused laser pulse is hard to self-focusing in the plasma at density of about $5.6\times10^{18} cm^{-3}$ corresponding $P_c = 52$ TW, so an electron beam with multiple spots and low energy is generated, as results shown in Figs. 2(a), 2(e); When plasma density is increased to $1.12\times10^{19} cm^{-3}$ ($P_c = 26$ TW), the matched laser $w_0 \approx R \approx 6 \mu m$ during self-focusing and $a_0 \approx 3$ is higher than the intensity of ionization threshold of nitrogen inner shell electrons\cite{ref36, ref37}. The electron beam charge is increased prominently to about 5 nC, far exceeding the charge scaling law of nonlinear bubble regime\cite{ref38} (the detail mechanism to be discussed later); With the increase of plasma density to $3.68\times10^{19} cm^{-3}$ ($P_c = 8$ TW), electron beam charge could be up to 20 nC and the divergence angle is just $\sim6^\circ$ [Figs. 2(c), 2(f)]. However, further increase plasma density, all of electron beam parameters deteriorate [Figs. 2(d-f)], due to the aggravated dephasing of electrons and the faster depletion of laser pulse in higher density plasma. It is worthy to mention that the optimal energy conversion efficiency from laser to collimated electron beam could be up to 12.4\% for $E_k > 1$ MeV [Fig. 2(g)], when the small focal spot laser pulse is matched by a suitable density plasma. Moreover, the beam energy is mainly located in the range of 5$\sim$30 MeV, which is ideal to drive $\gamma$-rays for stimulating photonuclear reactions in the giant dipole resonance region\cite{ref39}.

\begin{figure}[!h]
 \centering
	\includegraphics[width=0.5\textwidth]{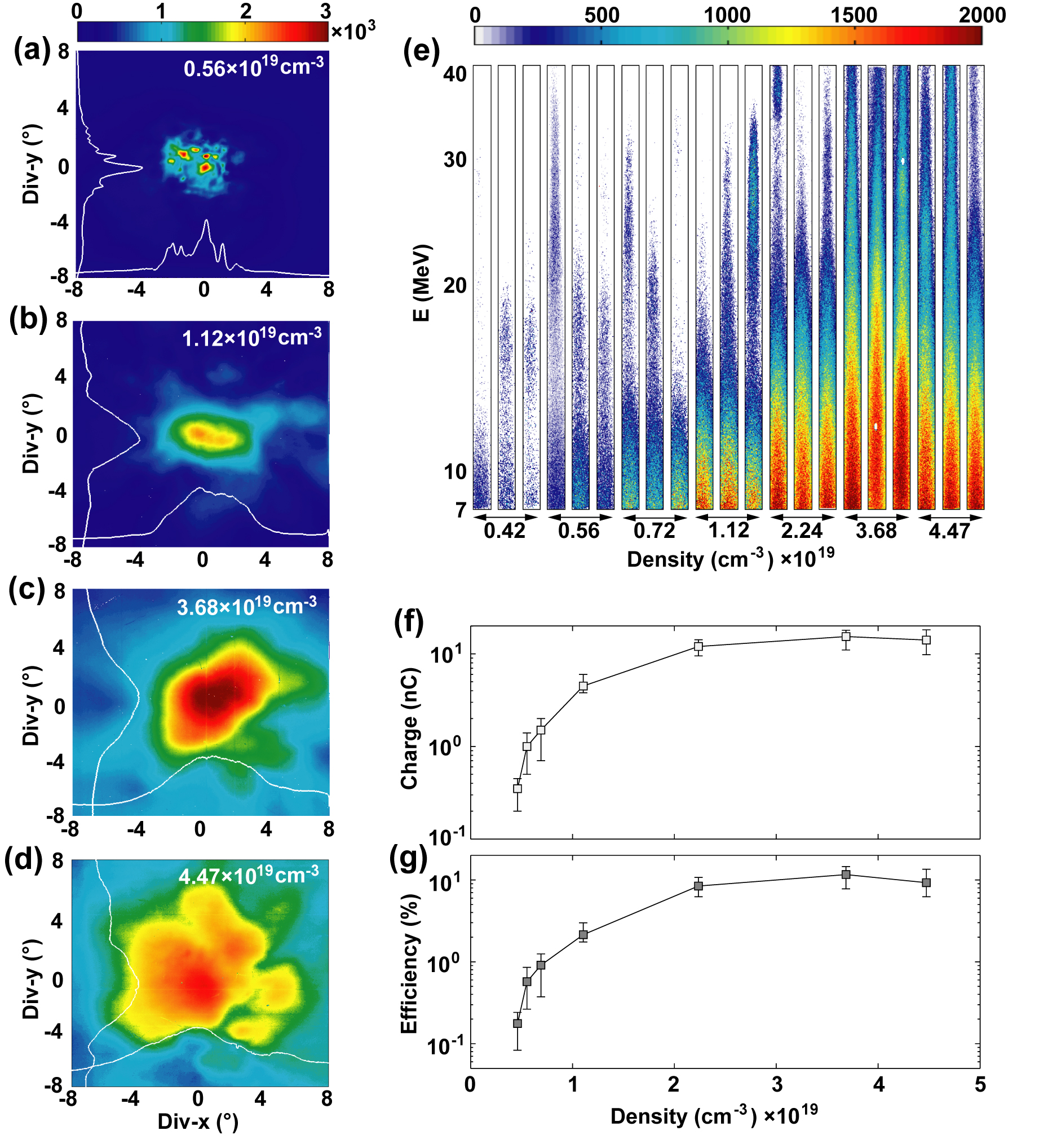}
 \caption{Experimental results of electron beam for different density plasma. (a-d) Electron beam angular distributions at plasma densities of 0.56, 1.12, 3.68 and $4.47\times10^{19} cm^{-3}$ respectively. (e) Electron beam energy distributions at different plasma densities. (f) and (g) is the corresponding total charge ($E >1$ MeV) and energy conversion efficiency of laser to electron beam respectively. }
\end{figure}

\begin{figure}[!h]
 \centering
	\includegraphics[width=0.5\textwidth]{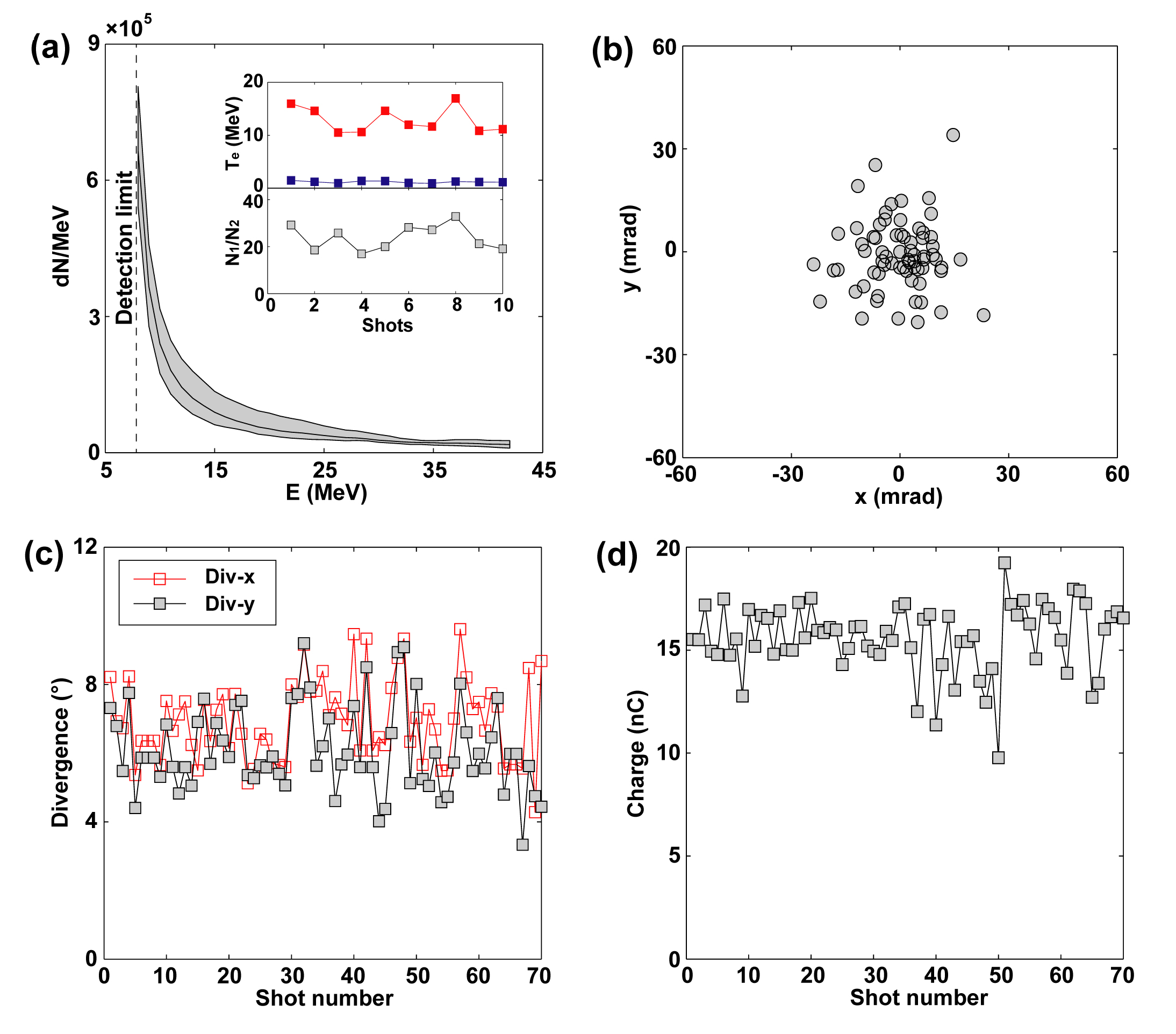}
 \caption{Experimental electron beam stabilities. (a) Electron beam energy spectrum of continuous 10 shots. The inset shows the temperatures of double temperature fitting and the ratio of the total number of electrons in the two temperatures. (b) Electron beam pointing of continuous 70 shots. (c) Electron beam divergence angle. (d) Electron beam charge.}
\end{figure}

\paragraph{Electron beam stabilities.}
To better demonstrate the characteristics of this optimized high efficiency laser plasma accelerator, electron beam stabilities, including energy spectrum, pointing, divergence angle and charge, are summarized as follow. The electron beam energy spectrums of continuous 10 shots are shown in Fig. 3(a). It is fitted by using double temperature $T_{e1} =1.19\pm0.19$ MeV, $T_{e2} =12.88\pm2.41$ MeV, and the ratio of the total number of electrons in the two temperatures is $N_1/N_2 =23.90\pm5.42$, here the error bars represent the standard deviation. The electron beam pointing stability of continuous 70 shots detected by DRZ is shown in Fig. 3(b), the standard deviation at x and y direction is 4.14 mrad and 5.09 mrad respectively. The corresponding divergence angle at x and y direction is $7.0^\circ\pm1.2^\circ$ and $6.1^\circ \pm1.3^\circ$ respectively [Fig. 3(c)]. Moreover, the electron beam charges are shown in Fig. 3(d), the average charge is 15.59$\pm$1.68 nC.

\section{\uppercase\expandafter{\romannumeral4}. SIMULATION AND ANALYSIS}

\paragraph{Particle in Cell Simulation.}
The two-dimensional particle in cell simulations were carried out using the EPOCH code\cite{ref40}. The simulation box size is $120\times120 \mu m^2$ with 4800$\times$1200 cells in the x and y directions. The simulation box propagates along the x-axis at the speed of light. Ionization is modelled with the ADK rates, one macro-particle per cell is used as nitrogen atom. The neutral nitrogen longitudinal profile has a 100 $\mu m$ up-ramp followed by a 1.2 mm long plateau with a uniform density, then followed by a 100 $\mu m$ down-ramp. The p-polarized laser pulse propagates along the x-direction, and the laser-focusing plane is located at the middle of up-ramp. The laser pulse has a Gaussian transversal profile with $w_0 = 3.8 \mu m$ and a Gaussian longitudinal envelope with pulse duration of 45 fs. The normalized vector potential $a_0 = 4.5$.

\begin{figure*}[!t]
 \centering
	\includegraphics[width=0.95\textwidth]{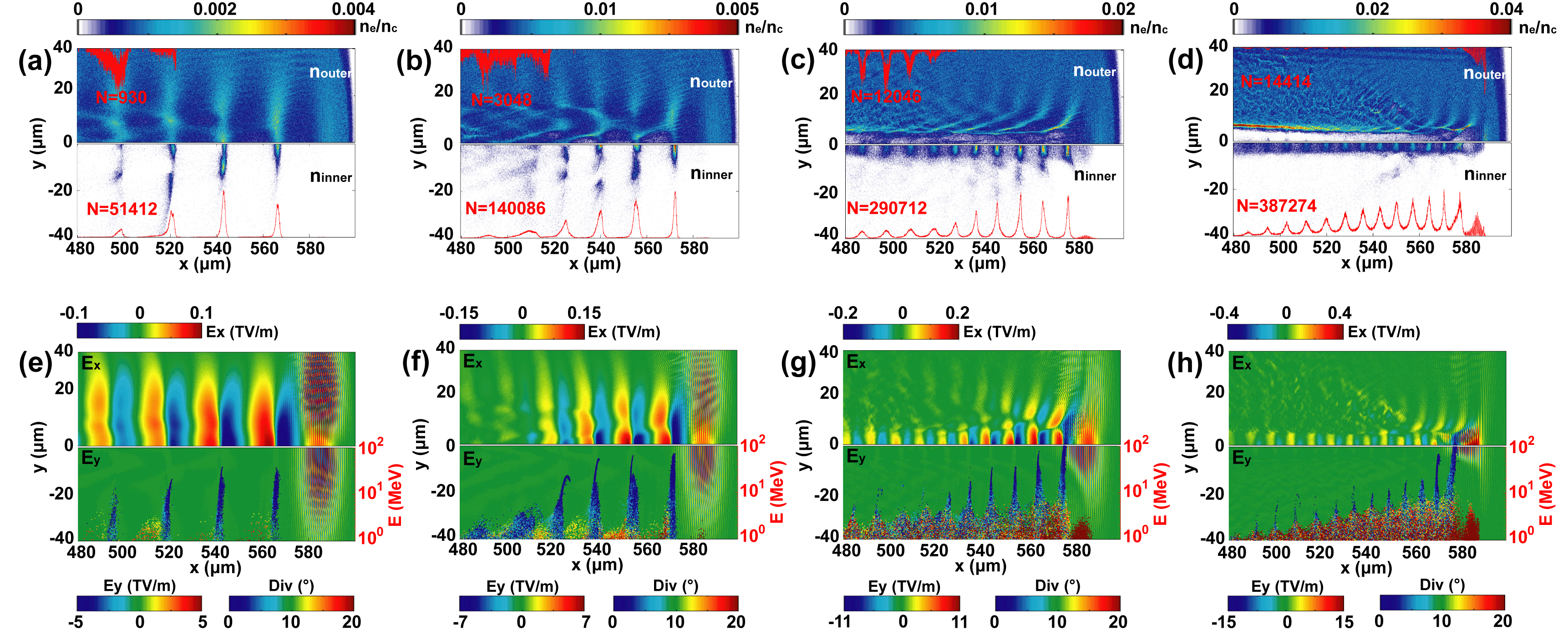}
 \caption{Small focal spot laser wakefield acceleration in pure nitrogen gas with different densities. (a-d) Plasma electron density distributions, where the upper/down represents the outer/inner shell electrons of nitrogen atom respectively, the red line (peak normalized) is electron number longitudinal distribution for $E_k \geq 1$ MeV and the number is the corresponding total micro particles. (e-h) Electric-field distributions, where the upper/down represents the $E_x/E_y$ respectively, the scattering points represent electron phase-space ($x-E_k$) distribution and the color represents electron divergence angle. (a, e) correspond to nitrogen gas density $4\times10^{17} cm^{-3}$; (b, f) $6\times10^{17} cm^{-3}$; (c, g) $1\times10^{18} cm^{-3}$; (d, h) $2.2\times10^{18} cm^{-3}$.}
\end{figure*}

\paragraph{Density matching condition for LWFA.}
To better understand the density matching condition for the generation of large charge, energetic and collimated electron beam, we studied the gas density influences on the small focal spot laser wakefield acceleration, as shown in Figs. 4. The drive laser pulse with power $\sim 30$ TW could not self-focusing in nitrogen gas with density $4\times10^{17} cm^{-3}$ (corresponding to fully ionized plasma density $5.6\times10^{18} cm^{-3}$ and $P_c =52$ TW), and its intensity decreases to $a_0\sim 1.25$ [Fig. 4(e)] which is lower than the ionization injection threshold of the $6^{th}$ electron of nitrogen atom ($a_0\sim 1.7$)\cite{ref37}. However, due to the initial laser $a_0\sim 4.5$ is much higher than that of the $7^{th}$ electron of nitrogen atom ($a_0\sim 1.9$), electrons would experience ionization injection [Fig. 4(a)] in the process of laser defocusing until $a_0$ less than the threshold of the $6^{th}$ electron, then electrons stop injecting as shown in Fig. 4(e). Significantly, many inner shell electrons are ionized and injected into successive multiple bubbles. When the gas density is increased to $6\times10^{17} cm^{-3}$, the corresponding $P_c\approx35$ TW is still higher than the power of drive laser, the laser pulse can not self-focusing but with slower defocus speed and keep $a_0\sim 1.7$ at the same distance [Fig. 4(f)], which results in more inner shell electrons ionization injection into bubbles [Fig. 4(b)]. If the gas densities continuously increase to $1\times10^{18} cm^{-3}$ and $2.2\times10^{18} cm^{-3}$ corresponding to $P_c \geq 21$ TW and 10 TW respectively, small focal spot laser would match to the plasma bubble during the self-focusing process, resulting in the laser intensity easily maintained above the threshold of the $7^{th}$ electron [Figs. 4(g, h)], and more inner shell electrons are injected into over ten bubbles to form high charge beam [Figs. 4(c, d)]. Although the higher density plasma is beneficial to increase the beam charge to a certain extent [Figs. 4(c, d)], the energy gain is lower due to that both the dephasing length $L_d\propto n_e^{-3/2}$ and the laser pump depletion length $L_p\propto n_e^{-1}$ are shorter\cite{ref34} in that cases [Figs. 4(g, h)]. That means we need to keep balance between electron charge and energy. In a word, with a suitable plasma density, the matched small focal spot laser pulse can maintain high intensity above the $7^{th}$ ionization threshold of nitrogen atom, and result in more inner shell electrons ionization inject into plasma wakefields.

\begin{figure*}[!t]
 \centering
	\includegraphics[width=0.9\textwidth]{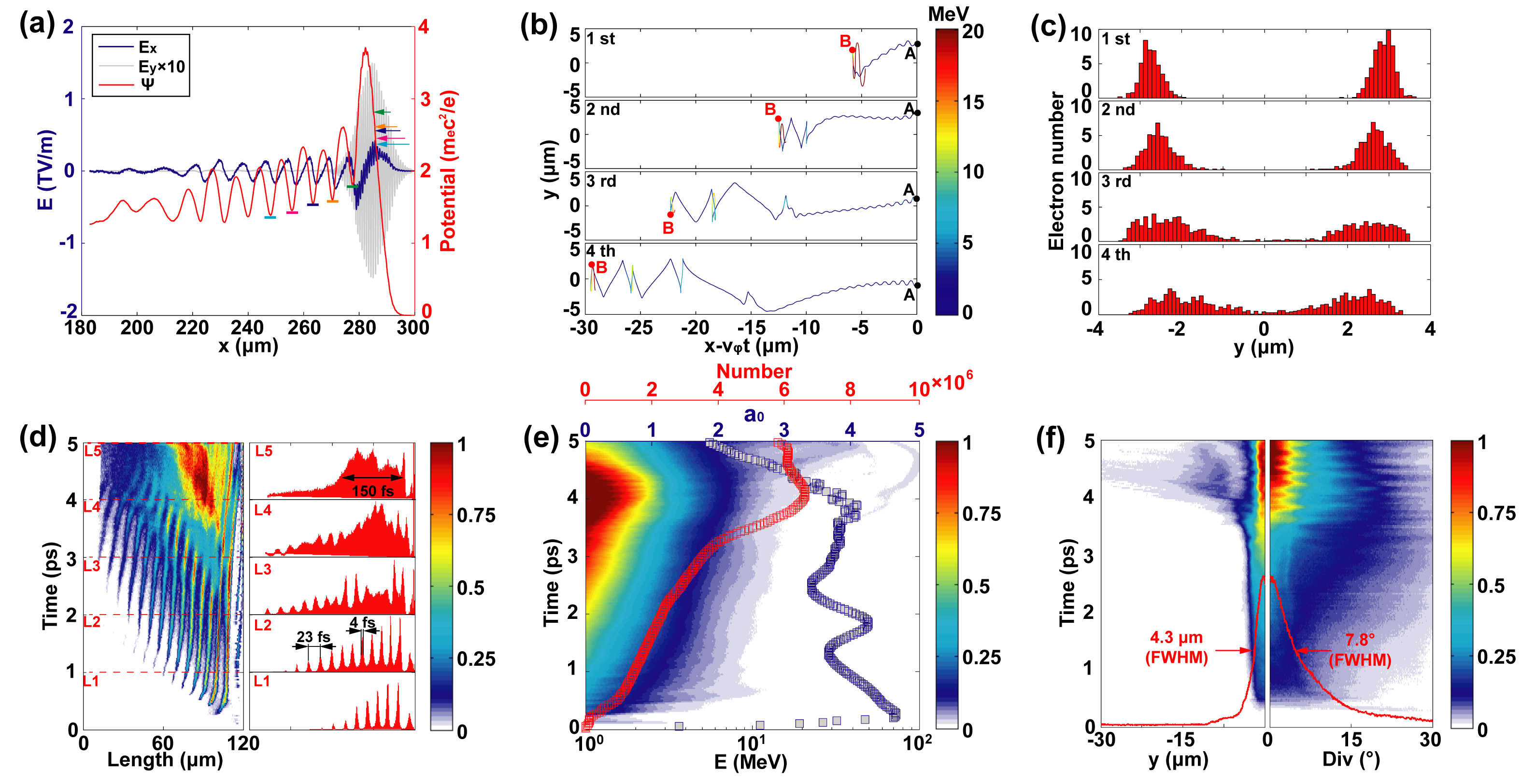}
 \caption{Analysis of multiple ionization injections and electron beam evolutions. (a) The axial electric-fields ($E_x$ bule line, $E_y$ gray line) of laser pulse at the moment of 1 ps and the corresponding electrostatic potential (red line) of $E_x$. (b) Four tracked electrons trajectories in the reference frame of drive laser pulse correspond to the $1^{st}$ $2^{nd}$ $3^{rd}$ and $4^{th}$ bubble respectively, in which the points A, B represent the position of ionization and injection respectively. (c) The transversal distribution of initial ionization position of electrons selected from $1^{st}$ $2^{nd}$ $3^{rd}$ and $4^{th}$ bubble respectively. (d) The evolution of electron beam longitudinal distribution. The right five figures correspond to the moments of 1, 2, 3, 4, 5 ps respectively. (e) The evolution of electrons energy distribution. The blue and red squares represent to the evolution of laser $a_0$ and total electron micro-particle number ($E_k \geq 1$ MeV) respectively. (f) The evolutions of transversal distribution and divergence angle of electron beam, in which the two red lines demonstrate the distributions of outgoing electron beam.}
\end{figure*}

In LWFA, the first wakefield or bubble in matched case, is usually more concerned due to the process of electron injection is more likely to occur inside\cite{ref41, ref42}. However, the number of electrons that can be loaded into a 3D nonlinear bubble is limited by the beam loading effect. According to M. Tzoufras$'$s theory\cite{ref38}, when the drive laser with a matched profile $k_pR\simeq 2\sqrt{a_0}$, the maximum charge of loaded electrons $\frac{Q_s}{1 nC}\simeq 0.047\sqrt{\frac{10^{16}cm^{-3}}{n_e}}(2\sqrt{a_0})^4 \frac{m_ec\omega_{pe}}{eE_s}$. According to the above conditions $a_0 \sim 4.5$, $n_e \sim 3.68\times10^{19} cm^{-3}$, and the acceleration field after reduction due to the beam loading effect $E_s \sim 0.1$ TV/m [Fig. 4(h)], the bubble can sustain the charge $Q_s$ could be $\sim$1.8 nC. However, in our case of small focal spot LWFA with high laser intensity, laser pulse can excite more wakefields and keep intensity exceeding the ionization threshold of nitrogen inner shell electrons. These electrons can be continuously ionized and injected into over ten bubbles [Fig. 4(g)], which results in the beam charge much higher than the limit of beam loading effect of a single bubble.

\paragraph{Multiple ionization injections and electron beam evolutions.}
To deeply understand the process of multiple ionization injection, the injection condition for the small focal spot LWFA in the nitrogen gas are presented [Figs. 5(a-c)]. Although the laser pulse would evolute due to the effect of self-focusing and etching in plasma, the plasma wake evolution is insignificant during propagating tens of microns. Thus, the electrostatic potential $\Psi$ of plasma wake can be used to qualitatively discuss the ionization injection conditions\cite{ref37}. Actually, electrons born at the peak of the wake potential ($E_z = 0$) will experience the largest $|\Delta \Psi_{max}|$, if $\Delta \Psi_{max}<-1$  then electrons will be ionized earlier in the wake and can be trapped\cite{ref36}. The maximum wake potential can be approximated as $\Psi_{max}\approx \frac{(k_pR)^2}{4}\approx a_0$, so the higher $a_0$ is propitious to trap electrons. By contrary, electrons born on pulse rising edge as shown blue arrow in Fig. 5(a), they would not experience a sufficient potential to be trapped in the first bubble. Fortunately, the potential in the subsequent wakefield is lower than the previous one due to the nonlinear blowout regime\cite{ref10,ref36}, resulting in these electrons could be trapped by the subsequent wakefield. In practice, if the inner shell electrons are ionized transversely far away from the axis of y=0, they will be longitudinally close to the peak of pulse envelope, and then they would experience sufficient potential to be trapped by the first bubble as shown in “1 st” of Fig. 5(b). However, when the electrons are ionized close to the axis of y=0, they will be born at the front edge of laser pulse. These electrons would be trapped by more backward wakefields, for example, as shown in “3 rd”, “4 th” of Fig. 5(b). Figure 5(c) demonstrates more intuitively that inner shell electrons ionized transversely close to the y=0 axis are injected into more backward bubbles.

To systematically introduce the characteristics of electron beam, the beam length evolution in plasma channel is presented in Fig. 5(d). At the beginning of LWFA process, the electron beam longitudinal distribution has many modulation peaks with interval $\sim$ 23 fs and duration $\sim$ 4 fs. As electrons continuously inject into plasma bubbles, these electron bunches begin to combine after $\sim$ 3 ps, then the outgoing electron beam has a duration of $\sim$ 150 fs (in FWHM). Hence the estimated beam peak current is $\sim$ 100 kA, according to the beam average charge of $\sim$ 15 nC in our experiment. Because the limit of dephasing length  $L_d\approx \frac{2\omega_p^2R}{3\omega_0^2}\approx 120 \mu m$, the electron maximum energy gain\cite{ref34} $\Delta E\approx \frac{2}{3}m_ec^2 (\frac{\omega_0^2}{\omega_p^2})^2a_0\approx 100$ MeV as shown in Fig. 4(h).Because beam loading effect reduces the acceleration electric-field, the energies of most electrons are located in the range of MeV to 30 MeV as shown in Fig. 5(e), which agree well with the experiment results. Due to the existence of self-generated confinement electromagnetic fields in the plasma bubbles\cite{ref42, ref43}, the electron beam has small source size (4.3 $\mu m$) and divergence angle ($7.8^\circ$) as shown in Fig. 5(f).

\begin{figure*}[!t]
	\centering
	\includegraphics[width=0.68\textwidth]{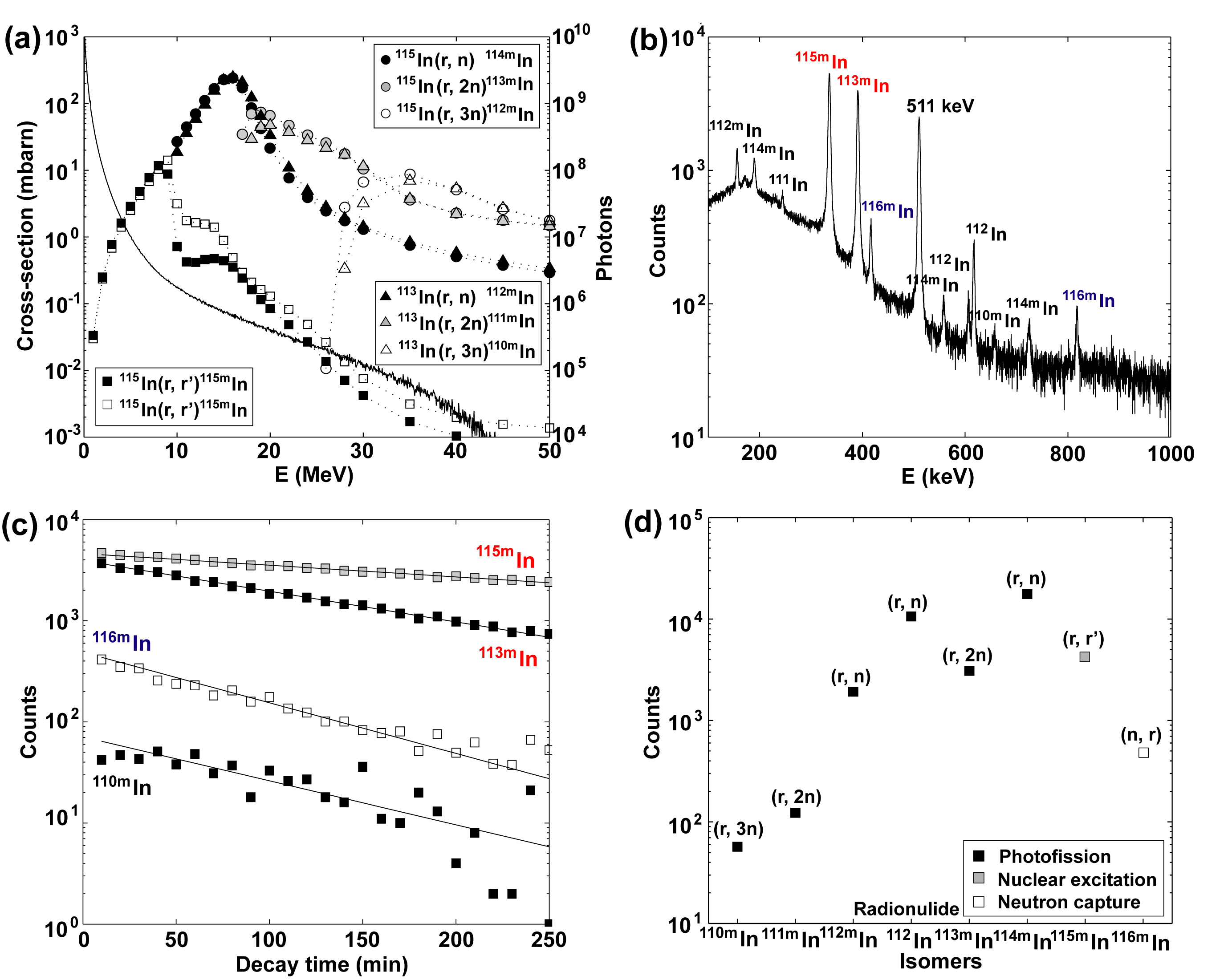}
	\caption{Results of isomers excitation. (a) The bremsstrahlung spectrum from W converter simulated by Geant4 code (solid line), and the cross-sections of photonuclear reactions from TENDL data library\cite{TENDL}. (b) A typical decay spectrum of In target measured by high pure germanium detector for 250 mins. (c) The experimental decay time spectrums. (d) The single shot yields of isomers and radionuclides produced by different nuclear reactions.}
\end{figure*}

This kind of tens MeV, large charge, collimated, short duration electron beam has great potential for driving an ultra-high flux photofission neutron source, an ultra-brightness gamma ray radiation source, or exciting and accumulating large amounts of isomers within ultra-short duration and tiny spatial size.

\section{\uppercase\expandafter{\romannumeral5}. ISOMER EXCITATION RESULTS}

Nuclear isomers have a broad range of potential applications\cite{ref44}. For example, new energy-storage materials\cite{ref45, ref46, ref47, ref48}, medical isotopes\cite{ref49, ref50, ref51}, nuclear clocks\cite{ref52, ref53}, or nuclear gamma-ray lasers\cite{ref54, ref55}. Especially for nuclear $\gamma$-ray lasers whose lifetimes of excited states are needed as short as nanosecond or even shorter\cite{ref55, ref56}, due to the relatively large emission linewidth of Doppler broadening, it is a big challenge for traditional accelerators or reactors to pump nuclei to these excited states efficiently\cite{ref57, ref58}. For example, considering a conventional commercial electron accelerator ($E = 5$ MeV, $I = 2$ mA, 500 Hz and duration of 15 $\mu s$) shooting on the same converter and target, the peak efficiency of $(\gamma,\gamma')$ reaction can be estimated to be $\sim 2\times10^7$ p/s. Even, for a traditional high energy electron accelerator (e.g., BEPC-\uppercase\expandafter{\romannumeral2} in China), the estimated peak efficiency could be $\sim10^{10}$ p/s for photonuclear reaction $(\gamma,n)$. In contrast, the electron beam from LWFA has ultra-high peak current (hundred kA), it has great potential to improve the excitation efficiency of nuclear isomers.

For production of nuclear isomers via photonuclear reactions driven by electron bremsstrahlung radiation, it is efficient for bremsstrahlung photon energy between 5 MeV and 30 MeV, e.g., $\gamma$-ray with energy $5\sim 10$ MeV is propitious to drive photonuclear reaction ($\gamma$, $\gamma'$) and excite nuclei\cite{ref59}, $10\sim30$ MeV $\gamma$-rays are ideal to stimulate giant dipole resonance for photofission reactions ($\gamma$, xn) and producing fast neutrons\cite{ref60} or nuclear isomers\cite{ref58}. However, for photon energy beyond 30 MeV, the reaction is not effective anymore, as well as the bresstrahlung photon yield\cite{ref61}.

To experimentally realize efficient photonuclear reactions in In target, a 1 mm thickness W was set close to the nozzle serving as a convertor for high brightness $\gamma$-ray radiation, and the radiation spectrum [Fig. 6(a)] was simulated by utilizing the above experimental electron beam parameters, the total photon number could be up to $1\times10^{11}$ for $E_p > 1$ MeV. According to the cross-sections of photonuclear reactions of In, the reactions of $^{*}$In$(\gamma,\gamma')^{*m}$In, $^{*}$In$(\gamma,n)^{*m}$In,  $^{*}$In$(\gamma,2n)^{*m}$In and $^{*}$In$(\gamma,3n)^{*m}$In can take place. In this experiment, we utilized electron beam with 0.025 Hz repetition rate to bombard the W+In target for isomer production. After one hour shooting, we took In target out of vacuum chamber, and measured its decay radiation. The decay radiation spectrum is shown in Fig. 6(b), and the different peaks from isomers or radionuclides are marked according to specialized radiation energy referring from NNDC database\cite{NNDCdatabase}. Four isomers decay time spectrums with several hours half-life are shown in Fig. 6(c), and their fitting half-lives agree quite well with the data of NNDC database.

The yield of $^{115m}$In from $^{115}$In$(\gamma,\gamma')^{115m}$In is about $4.16\times10^3$ for single shot [Fig. 6(d)]. However, the yield of $^{114m}$In from $^{115}$In$(\gamma,n)^{114m}$In could be up to $1.76\times10^4$ due to the higher photofission cross-section. The $^{113m}$In is contributed by $^{113}$In$(\gamma,\gamma')^{113m}$In and $^{115}$In$(\gamma,2n)^{113m}$In with the total yield of $\sim 3.09\times10^3$. The $^{111m}$In excited by $^{113}$In$(\gamma,2n)^{111m}$In has the yield of $\sim 1.23\times10^2$. Therefore, according to the abundance ratio of $^{113}$In and $^{115}$In, the yield of $^{113m}$In excited from $^{115}$In$(\gamma,2n)^{113m}$In is estimated about $2.74\times10^3$. In addition, the $^{116m}$In has a yield of $\sim 4.85\times10^2$ from $^{115}$In$(n,\gamma)^{116m}$In which is induced by photofission neutrons. Because the gamma ray beam duration is close to the electron beam, the time of $\gamma$-ray pass through 3 mm thickness In target is about 10 ps, which can be regarded as the exciting duration time. Therefore, the peak excitation efficiency of isomers from photonuclear reactions are estimated about $4.16\times10^{14}$ p/s $(\gamma,\gamma')$, $1.76\times10^{15}$ p/s $(\gamma,n)$ and $2.74\times10^{14}$ p/s $(\gamma,2n)$ respectively. Limited by the vacuum pump efficiency and laser repetition rate in this experiment, the average excitation efficiency of isomer is less than $5\times10^2$ p/s. However, average efficiency $10^7$ p/s can be realized by utilizing a more powerful pump set and a 100 Hz hundred-TW laser facility which is available right now. This pumping method will also greatly benefit to the production of medical isotopes and nuclear batteries on tabletop.

\section{\uppercase\expandafter{\romannumeral6}. CONCLUSION}
In summary, we have presented a novel efficient electron injection method in laser plasma wakefield acceleration. The inner shell electrons of nitrogen atom would be continuously ionized injection into multiple bubbles, when the small focal spot intense laser pulse is matched in a suitable higher density nitrogen gas. A hundred kilo-ampere electron beam has been generated with 12.6\% energy conversion efficiency from the driving laser. The electron beam has average charge of $\sim$15 nC, divergence angle of $\sim6^\circ$ and suitable energies for photonuclear reactions. By utilizing this high current electron beam to drive the excitation of nuclear isomers via reactions of $(\gamma,\gamma')$, $(\gamma,xn)$, an ultra-high isomer pumping peak efficiency $\sim1.76\times10^{15}$ p/s has been realized in In target, which is at least five orders of magnitude higher than using traditional electron accelerators. This efficient and easily accessible production method of exciting nuclear isomers during short time could be greatly beneficial for the study of nuclear transition mechanisms and nuclear gamma ray lasers.

\section{ACKNOWLEDGEMENTS}
This work is supported by the Science Challenge Project (TZ2018005), the National Nature Science Foundation of China (11875191, 11805266, 11991073, 11890710, 11721404), the Strategic Priority Research Program of the CAS (XDB1602, XDA01020304), the National Key R\&D Program of China (2017YFA0403301).


\end{document}